\documentstyle[12pt]{article}

\def\a{\alpha}
\def\p{\psi}
\def\e{\epsilon}

\def\Z{{\bf Z}}

\def\bar{\overline}
\def\hat{\widehat}

\begin{document}

\renewcommand{\theequation}{\thesection.\arabic{equation}}
\newcommand{\beq}{\begin{equation}}
\newcommand{\eeq}{\end{equation}}
\newcommand{\bea}{\begin{eqnarray}}
\newcommand{\eea}{\end{eqnarray}}
\newcommand{\bq}{\begin{equation}}
\newcommand{\eq}[1]{\label{#1}\end{equation}}
\newcommand{\ber}{\begin{eqnarray}}
\newcommand{\eer}[1]{\label{#1}\end{eqnarray}}

\newcommand{\be}{\begin{equation}}
\newcommand{\ee}{\end{equation}}

\newcommand{\ba}{\begin{eqnarray}}
\newcommand{\ea}{\end{eqnarray}}

\begin{center}
        April, 1994
                                \hfill    IASSNS-HEP-94/30\\
                                \hfill    RI-157-93\\
                                \hfill    hep-th/9404184

\vskip .5in

{\large \bf Mirror Symmetry as a Gauge Symmetry}
\vskip .5in

{\bf Amit Giveon} \footnotemark \\

\footnotetext{e-mail address: giveon@vms.huji.ac.il}

\vskip .1in

{\em Racah Institute of Physics, The Hebrew University\\
  Jerusalem, 91904, ISRAEL} \\

\vskip .15in

 and

\vskip .15in

{\bf Edward Witten} \footnotemark \\

\footnotetext{e-mail address: witten@sns.ias.edu}

\vskip .1in

{\em School of Natural Sciences \\
Institute for Advanced Study \\
Olden Lane, Princeton N.J., USA}\\
\vskip .1in
\end{center}
\vskip .4in

\begin{center} {\bf ABSTRACT } \end{center}
\begin{quotation}\noindent

It is shown that in string theory mirror duality is a gauge symmetry
(a Weyl transformation)
in the moduli space of $N=2$ backgrounds on group manifolds,
and we conjecture on the possible generalization to
other backgrounds, such as Calabi-Yau manifolds.

\end{quotation}
\vfill
\eject

\section{Introduction}
\setcounter{equation}{0}

Target-space dualities in string theory are symmetries relating backgrounds
with {\em different geometries} that correspond to the same 2-$d$ Conformal
Field Theory (CFT) \cite{duality}. The simplest example is the $R\to 1/R$
circle duality \cite{KY},
that relates a circle of radius $R$ to a circle of radius $1/R$.

This duality is a gauge symmetry in string theory
in the following sense \cite{DHS}.
At the self-dual point, $R=1$, there is an enhanced $SU(2)_L\times SU(2)_R$
affine symmetry. One can deform away from the $R=1$ point by
adding a current-current truly marginal operator, $J\bar{J}$.
There is a Weyl rotation in $SU(2)_L$ that takes $J\to -J$, and therefore,
this transformation is a symmetry of the {\em self-dual} point.
However, this transformation relates the conformal
deformation $\epsilon J\bar{J}$ to $-\epsilon J\bar{J}$, and on the full
modulus line of circle compactifications, this transformation
corresponds to the $R\to 1/R$ duality.

In the target space effective action we have the following picture (for a
review, see e.g. \cite{GSW}).
The worldsheet couplings to operators, perturbing a given 2-$d$ action,
become target space fields.
There is an $SU(2)_L\times SU(2)_R$ gauge symmetry when the scalar fields
get VEVs that correspond to the $R=1$ point.
This gauge symmetry is spontaneously broken to $U(1)_L\times U(1)_R$ when
one changes the VEVs of scalar fields. There is a residual $\Z_2$ gauge
transformation in the spontaneously broken gauge group that relates the VEV
corresponding to radius-$R$ compactification to the VEV corresponding to
radius-$1/R$ compactification.
It is in this sense that the $R\to 1/R$ duality is a gauge symmetry in
string theory.

The interpretation of target-space dualities as gauge symmetries was
generalized to the duality group $O(d,d,\Z)$ of toroidal backgrounds
\cite{GMR1}. Moreover, in ref. \cite{GK} it was shown that there is a
duality in the moduli space of $J\bar{J}$ deformations of  $G_k$ WZW
models. This duality is a gauge symmetry in the same sense described above.
It is called an `axial-vector duality' for reasons that will be clear soon
\cite{GK}, and it relates {\em curved} backgrounds with different
geometries, and even with {\em different topologies}.

The relation of target space dualities to gauge symmetries shows that they
are {\em exact} symmetries in string theory (to all orders and interactions).

In this note we will describe a particular target-space duality --
{\em mirror symmetry} \cite{mir} --
in the moduli space of $N=2$ backgrounds on a group manifold $G$.
Moreover, mirror duality will be related to a gauge symmetry in string
theory.

The structure of the paper is as follows: In section 2, we begin with an
$N=2$ affine construction on a group $G$, and in section 3, we consider the
$N=2$ construction on $SU(2)\times U(1)$. In section 4, we describe the mirror
transformation, and in section 5, we discuss mirror duality in the moduli
space of $N=2$ models derived from
$SU(2)\times U(1)$ (or $SL(2)\times U(1)$). In section 6, we
discuss mirror duality as a gauge symmetry in the moduli space of $N=2$
models on general groups $G$, and in section 7, we present the $SU(3)$
example. Finally, in section 8, we conjecture that mirror duality is a
gauge symmetry in string theory, also for Calabi-Yau compactifications.

\section{$N=2$ Affine Construction on a group $G$ }
\setcounter{equation}{0}

It is known that any even dimensional group allows an $N=2$ super affine
symmetry \cite{SSTV}. Following ref. \cite{KS},
we generate the affine $N=2$ algebra on a group
$G$ at level $k$. It is sufficient to describe the left-handed part.
Let us present currents $j^a(z)$ and fermions $\p^a(z)$ in the adjoint
of $G$ that satisfy the operator product expansion (OPE)
\be
j^a(z)j^b(w)={\hat{k}\delta^{ab} \over (z-w)^2}+{if_{abc}j^c(w) \over
z-w}+...,
\label{jajb}
\ee
\be
\p^a(z)\p^b(w)={{k \over 2}\delta^{ab} \over z-w}, \qquad
j^a(z)\p^b(w)=0+....
\ee
Here $f_{abc}$ are the structure constants of the Lie algebra $G$,
$\hat{k}\equiv k-C_2(G)$, $f_{apq}f_{bpq}=\delta_{ab}C_2(G)$, and dots
stand for
non-singular terms in the OPE \footnote{As indices are raised and lowered
by $\delta_{ab}$ we will not be careful about upper and lower indices. Any
repetition of indices means a summation. The discussion can be carried out
for a general bilinear form $\eta$ replacing $\delta$.}.
The sigma-model which corresponds to this theory is the level $k$
$N=1$ WZW Lagrangian on a group $G$,
\be
S[\hat{g}]={k \over 2\pi}\int d^2zd^2\theta {\rm Tr} \left(
D\hat{g}^{-1}\bar{D}\hat{g}-i\int
dt[\hat{g}^{-1}D\hat{g},\hat{g}^{-1}\partial_t\hat{g}]
\hat{g}^{-1}\bar{D}\hat{g}\right),
\ee
where
\be
\hat{g}(z,\bar{z},\theta,\bar{\theta})=e^{T_aX^a}, \qquad
X^a=x^a+\theta {\psi^a \over k} + \bar{\theta}{\bar{\psi}^a \over k}
+\bar{\theta}\theta F^a, \qquad D={\partial \over
\partial\theta}+\theta{\partial \over \partial z}.
\ee
The chiral $N=1$ supercurrents are
\be
\hat{J}^a=k{\rm Tr}(T^a D\hat{g}\hat{g}^{-1})=\psi^a+\theta\left(j^a-{i\over
k}f^a_{bc}\psi^b\psi^c\right),
\label{hatJ}
\ee
where the currents $j^a$ in (\ref{jajb},\ref{hatJ}) are given by
\be
j^a=k{\rm Tr}(T^a \partial g g^{-1}), \qquad  g=e^{T_ax^a}.
\label{jg}
\ee
The central charge of this model is
\be
c={\hat{k}{\rm dim}G \over \hat{k}+C_2}+{1\over 2}{\rm dim}G.
\ee

For the $N=2$ superconformal algebra (SCA),
in addition to the stress-tensor and $N=1$ supercurrent,
\be
T(z)={1\over k}(j^a j^a - \p^a\partial \p^a),
\ee
\be
G^0(z)={2\over k}(\p^a j^a-{i \over 3k}f_{abc}\p^a\p^b\p^c),
\ee
we need another $N=1$ supercurrent which we write as
\be
G^1(z)={2\over k}(h_{ab}\p^a j^b-{i \over 3k}S_{abc}\p^a\p^b\p^c).
\ee
We define $G^{\pm}$ by
\be
G^0\equiv {1 \over \sqrt{2}}(G^+ + G^-), \qquad
G^1\equiv {1 \over \sqrt{2}i}(G^+ - G^-) .
\ee
The necessary and sufficient conditions for achieving $N=2$ SCA are
\be
h_{ab}=-h_{ba}, \qquad h_{ac}h_{cb}=-\delta_{ab},
\label{con1}
\ee
\be
f_{abc}=h_{ap}h_{bq}f_{pqc}+h_{bp}h_{cq}f_{pqa}+h_{cp}h_{aq}f_{pqb},
\label{con2}
\ee
\be
S_{abc}=h_{ap}h_{bq}h_{cr}f_{pqr}.
\label{con3}
\ee
When these conditions are satisfied, an $N=2$ SCA is generated by
$T(z),G^+(z),G^-(z), J(z)$ (see for example \cite{KS}),
where the $U(1)$ current, $J$, is determined from the explicit OPEs:
\be
J=h_{ab}\left[{i\over k}\p^a\p^b+{1\over k}f^{ab}_c(j^c-{i\over
k}f^c_{de}\p^d\p^e)\right].
\label{J}
\ee

The condition (\ref{con1}) means that $h_{ab}$ is an (almost)
complex structure. To see the meaning of conditions
(\ref{con2}),(\ref{con3}), let us introduce the projection operators
\be
(P_{\pm})_{ab}={1\over 2}\left(\delta_{ab}\pm {1\over i}h_{ab}\right),
\label{P}
\ee
and split the set of the Lie  algebra generators
$T=\{T^a | [T^a,T^b]=if_{abc}T^c \}$ into two sets $T_+$ and $T_-$:
\be
T_{\pm}=\{T_{\pm}^a | T_{\pm}^a=(P_{\pm})_{ab}T^b\}.
\ee
Then, by a straightforward calculation one finds that
(\ref{con2}),(\ref{con3}) are equivalent to the conditions
\be
\left[T_{\pm}^a,T_{\pm}^b\right]={i \over 2}(f_{abc}\pm iS_{abc})T_{\pm}^c,
\ee
which can be written schematically as
\be
\left[T_+,T_+\right]\subset T_+, \qquad
\left[T_-,T_-\right]\subset T_- .
\ee
We may thus summarize the result as follows \cite{KS}:

\vskip 0.1in

\noindent
{\bf Theorem}:

\noindent
Let $T$ be the complexified Lie algebra of $G$. Then the model
$G$ has an $N=2$ structure for every direct sum decomposition $T=T_+\oplus
T_-$ ( ${\rm dim}\, T_+={\rm dim}\, T_-$ ) such that $T_+$ and $T_-$ separately
form a closed Lie algebra, and $T_- = \bar{T}_+$.

For the applications of this result, one must bear in mind the following.
Our discussion so far has been purely algebraic.  In a geometric context
of WZW models, one has both left and right-moving current algebra, coming
from the left and right action of $G$ on itself.  Accordingly, two
copies of $T$ appear, say $T_L$ and $T_R$ -- the generators of the left
and right action of $G$, which we will call $G_L$ and $G_R$.
In constructing an $N=2$ structure -- by
which we mean a structure with $(2,2)$ supersymmetry -- with target space
$G$, the above theorem must be used twice, once for left-movers and
once for right-movers.  Accordingly, one actually picks two complex
structures on $T$, a left-moving one and a right-moving one.

\section{$N=2$ Construction on $SU(2)\times U(1)$ }
\setcounter{equation}{0}

Let $T=\{T_1,T_2,T_3,T_0\}$, where $\{T_i,\,
 | i=1,2,3\}$ are the generators of
the $SU(2)$ Lie algebra,  $\left[T_i,T_j\right]=i\epsilon_{ijk}T_k$,
and $T_0$ is the $U(1)$ generator.
A complex structure
\be
h_{ab}=\left(\matrix{\e & 0 \cr 0 & \e \cr}\right), \qquad
\e= \left(\matrix{0 & 1 \cr -1 & 0 \cr}\right) ,
\label{hsu2u1}
\ee
gives an $N=2$ SCA, as was shown in  \cite{SSTV,RSS}.
The proof can be done either by a straightforward check that the conditions
(\ref{con1}),(\ref{con2}) are satisfied ((\ref{con3}) defines $S_{abc}$ in
terms of $h_{ab}$), or by showing that the structure described in
the theorem is maintained. Let us do the latter: the projection operators are
\be
P_{\pm}=\left(\matrix{I\mp i\e & 0 \cr 0 & I\mp i\e \cr}\right), \qquad
I= \left(\matrix{1 & 0 \cr 0 & 1 \cr}\right) ,
\ee
and therefore,
\ba
T_+ = \{T_+^1={1\over 2}(T_1-iT_2),\; T_+^2={1\over 2}(T_3-iT_0)\},
\nonumber\\
T_- = \{T_-^1={1\over 2}(T_1+iT_2),\; T_-^2={1\over 2}(T_3+iT_0)\}.
\ea
One finds that
\be
\left[T_+^1,T_+^2\right]={1\over 2}T_+^1, \qquad
\left[T_-^1,T_-^2\right]=-{1\over 2}T_-^1,
\ee
and therefore, $\left[T_+,T_+\right]\subset T_+, \;
\left[T_-,T_-\right]\subset T_- $.

\section{Mirror Transformation }
\setcounter{equation}{0}

For simplicity, we first describe the $SU(2)\times U(1)$ model.
Combining left-movers and right-movers we have an $N=2$ affine algebra on
$(SU(2)\times U(1))_L \times (SU(2)\times U(1))_R$. A mirror transformation,
$m$, is a transformation of $N=2$ CFT's that acts  as
\be
m\;\; : \;\; J\to -J, \qquad \bar{J}\to \bar{J},
\ee
where $J$ ( $\bar{J}$ ) is the left- (right-) handed $N=2$ $U(1)$ current.
{}From (\ref{J}) it follows that in the present context
$m$ acts on the left-handed
complex structure as
\be
m(h_{ab})=-h_{ab},
\label{htoh}
\ee
while commuting with the right-handed one.

We now arrive to a key point.  If the left and right moving complex
structures are as described above, then a Weyl rotation in the group
$SU(2)_L$ has the right properties to be interpreted as a mirror symmetry.
In the realization of the $SU(2)\times U(1)$ model as a WZW model, the
field $g$ in $SU(2)\times U(1)$ transforms to $mg$.
We pick $m$ to a $\pi$-rotation around the 1-axis,
\be
m=\left(\matrix{ 0 & i\cr
                i & 0 \cr}\right),
\ee
acting on the Lie algebra as
\be
m(T_1,T_2,T_3,T_0)=(T_1,-T_2,-T_3,T_0).
\ee
Thus, $m$ interchanges $T_+$ with $T_-$, and therefore, it takes the
left-handed complex structure to its minus.
Thus, the $SU(2)\times U(1)$ model, with the $N=2 $ structure under
discussion, is equivalent to its own mirror, via the transformation $m$.

\section{Mirror Duality in the Moduli Space of $N=2$
$SU(2)\times U(1)$ (or $SL(2)\times U(1)$)}
\setcounter{equation}{0}

We now look for a current-current deformation, $W\bar{W}$, where
\ba
W=\sum_{a=0}^3 \alpha_a J^a, \qquad J^a={\rm Tr}W^a=
{\rm Tr}\left[T^a \left(k\partial g g^{-1}-{2\over k}T_b\psi^bT_c\psi^c\right)
\right],
\qquad g\in SU(2)\times U(1), \nonumber\\
\bar{W}=\sum_{a=0}^3 \beta_a \bar{J}^a, \qquad \bar{J}^a={\rm Tr}\bar{W}^a
= {\rm Tr}\left[T^a \left(k g^{-1} \bar{\partial}g
-{2\over k}T_b\bar{\psi}^bT_c\bar{\psi}^c\right)\right],
\ea
such that the chiral current $J^a$ (antichiral current) is an
upper component of the chiral $N=1$ supercurrent in (\ref{hatJ})
(antichiral supercurrent). When $J^a$ and $\bar{J}^a$ are in the Cartan
sub-algebra, the $W\bar{W}$ deformation preserves $N=2$ supersymmetry.
(This is explained
in section 6, in a more general case.)

The deformation $W\bar{W}$ is particularly interesting if $W=J^3$,
$\bar{W}=\bar{J}^3$. This deformation is {\em odd} under the mirror
symmetry $m$,
\be
m(W\bar{W})=-W\bar{W}.
\label{mww}
\ee
This is true as $m$  anticommutes with $W$ and commutes with $\bar{W}$:
\be
\{m,W\}=0, \qquad [m,\bar{W}]=0.
\ee
The first equality is true because $m$ is a rotation around the 1-axis while
$W=J^3$ is a rotation around the 3-axis. The second equality is trivially
true as $m$ acts purely in the left-handed sector.

The meaning of eq. (\ref{mww}) is that under the mirror transformation $m$,
the (infinitesimal) perturbation $\e W\bar{W}$ is related to
$-\e W\bar{W}$ (as $W\bar{W}$ is mirror odd). Therefore, {\em mirror
symmetry is a gauge transformation} ($m\in SU(2)_L$)
along the $W\bar{W}$ deformation line.

This deformation line was already studied in ref. \cite{GK} (although
for $N=0$ WZW models). The perturbation operator
\be
W\bar{W}=J^3\bar{J}^3=j^3\bar{j}^3+({\rm terms}\;\;{\rm  with}\;\; {\rm
worldsheet}\;\;{\rm  fermions})
\label{WWJJ}
\ee
deforms the $SU(2)$ WZW sigma-model, and generates
a one-parameter family of conformal sigma-models parametrized by
$0<R<\infty$ (we refer the reader to ref. \cite{GK} for details).
Together with the extra $U(1)$ and worldsheet fermions,
these sigma-models are $N=2$ backgrounds\footnote{
The terms in $J^3\bar{J}^3$ which depend on $\psi,\bar{\psi}$ must change
the quadratic and quartic fermionic terms in the Lagrangian, in a way
compatible with the worldsheet supersymmetry.}.
The mirror duality is nothing but the axial-vector duality of
\cite{GK}, which relates the model $R$ to the model $1/R$.
In particular, duality relates the two boundaries of the $R$-modulus
($R\to 0,\infty$) where the conformal sigma-models correspond to
$(SU(2)/U(1))_a\times U(1)\times U(1)_{\e\to 0}$ and
$(SU(2)/U(1))_v\times U(1)\times U(1)_{\e\to 0}$.
Here $U(1)_{\e\to 0}$ denotes a compact, free scalar field at the
limit when its compactification radius approaches 0, and
$a$ ( $v$ ) denotes the axially gauged (vectorially gauged) $SU(2)/U(1)$.
Therefore, mirror symmetry relates the axial Abelian coset to the
vector coset.
These two (equivalent) descriptions of
the parafermionic CFT are related by a $Z_k$ orbifolding \cite{GQ,EK}.

An alternative description of the models along the deformation line
(\ref{WWJJ}) is the sum of a parafermionic action and the action of a free
scalar field with radius $\sqrt{k}R$, up to a $Z_k$ orbifoldization which
couples the two \cite{GK,Yang}.
The orbifolding acts as a $Z_k$ twist of the parafermionic
theory and a simultaneous translation of the free scalar by
$2\pi(\sqrt{k}R)/k$. At the boundary $R\to\infty$ the twisted sectors
decouple, because a non-zero winding of the scalar field has infinite energy.
In the untwisted sector, every $Z_k$-eigenstate of the parafermion combines
with a continuum of the free scalar states to form  $Z_k$-invariant
states. Therefore, at $R\to\infty$ one gets the direct product of an
untwisted parafermion with a non-compact ($R\to\infty$) scalar and another
scalar field.
At the boundary $R\to 0$, since non-zero windings do not carry energy,
the $Z_k$ twist acts purely in the parafermionic sector. Thus, at $R\to 0$
one gets the direct product of a $Z_k$-orbifold of a parafermion with an
$R\to 0$ scalar and another scalar field.
In this description, mirror symmetry acts as a $Z_k$ orbifold on the
$N=2$ minimal model, and as a factorized duality \cite{duality}
on the two scalar fields.

The discussion above is even more interesting when $SU(2)$ is being
replaced by $SL(2)$.\footnote{We define the CFT corresponding to $SL(2)$
to be the one regularized by its Euclidean continuation, see \cite{GK}.}
The  $SL(2)\times U(1)$ model  has an $N=2$ structure,
and mirror duality is a
gauge symmetry as $m\in SL(2)$. The two boundaries (related to each other
by mirror transformation) correspond to
$(SL(2)/U(1))_a\times U(1)\times U(1)_{\e\to 0}$ and
$(SL(2)/U(1))_v\times U(1)\times U(1)_{\e\to 0}$.
The axial-vector duality in the $SL(2)/U(1)$ case relates backgrounds with
different geometries, and even {\em different topologies} (the
semi-infinite ``cigar'' and the infinite ``trumpet''); this is the
2-$d$ black-hole duality \cite{G}.

\section{Mirror Duality as a Gauge Symmetry in the Moduli Space of
$N=2$ $G$ Models}
\setcounter{equation}{0}

The discussion in the previous sections is not limited to the $SU(2)\times
U(1)$ ($SL(2)\times U(1)$) case, and can be extended to general groups, $G$,
that admit $N=2$. In fact, the theorem of section 2 can be applied
to any group with even rank, rank$\, G=2n$ \cite{SSTV}.
To do this, one picks a complex structure on the Cartan subalgebra,
that is, we split the generators of
the Cartan sub-algebra into two complex-conjugate sets $H_+$, $H_-$, such that
${\rm dim}\, H_+={\rm dim}\, H_-=n$, and set
\be
T_+=\{E_{\a_+},H_+\}, \qquad T_-=\{E_{\a_-},H_-\}.
\ee
Here $E_{\a_+}$ ($E_{\a_-}$) is the set of generators corresponding to
positive (negative) roots.
It is obvious that ${\rm dim}\, T_+={\rm dim}\, T_-$ ($={\rm dim}\, G/2$) and
$[T_+,T_+]\subset T_+,\; [T_-,T_-]\subset T_-$. Now, we define $h_{ab}$
(and therefore, the $N=2$ current $J$) in the basis $\{T_+,T_-\}$ to be
\be
h=\left(\matrix{iI & 0 \cr 0 & -iI \cr}\right),
\ee
namely,
\be
h(T_+)=iT_+, \qquad  h(T_-)=-iT_- .
\label{hTT}
\ee

A mirror transformation, $m$, should take
$h\to -h$, and from (\ref{hTT}) it follows
that it should interchange $T_+$ with $T_-$:
\be
m(T_+)=T_- , \qquad m(T_-)=T_+ .
\label{mTT}
\ee
Such a mirror symmetry can be realized as a symmetry of the $N=2$ model
iff $m$ is a Weyl rotation:
\be
m \;\; {\rm is\;\; mirror \;\; and \;\; gauge \;\; symmetry} \;\;
\Leftrightarrow  \;\; m(h)=-h, \;\; m\in G_L ,
\label{iff}
\ee
namely, when there is a Weyl rotation that takes $T_+\leftrightarrow T_-$.
When (\ref{iff}) is satisfied,
the $N=2$ WZW model on $G$ is {\em self-mirror}.
Moreover, when one allows for marginal deformations
as above, the mirror transformation acts non-trivially on the resulting
$N=2$ moduli space. (If $m$ is not a Weyl rotation, then this mirror
transformation is not
a symmetry of the given $N=2$ structure of the WZW model but maps
that structure to another one.)

Let us discuss $N=2$ preserving superconformal deformations, $W\bar{W}$,
of the $N=2$ $G$ model. By performing an Abelian duality (for a review,
see \cite{duality}),
one finds that a $G$ WZW model is equivalent to $[G/U(1)^r]\times U(1)^r$,
$r={\rm rank}\, G$
(up to an orbifolding by a finite discrete group) \cite{RV,EK}.
Any deformation of the $U(1)^r$ torus preserves
$N=2$. In the $G$ WZW model, such conformal perturbations correspond to
deforming the maximal torus, namely, to $W\bar{W}$ in the Cartan
sub-algebra $H$.
Therefore, any perturbation of the form $\e_{ij}H^i\bar{H}^j, \;\;
i,j=1,...,r, \;\; H^i,\bar{H}^j\in H$, preserves $N=2$.

Now, under mirror transformation, $m$, $H^i\to m(H^i)=m^i_{\;k}H^k$, and
therefore,
\be
m:\; \e_{ij}H^i\bar{H}^j\to (m^t\e)_{ij}H^i\bar{H}^j .
\ee
As a consequence,
the sigma-model backgrounds, corresponding to the deformations
$\e_{ij}$ and $(m^t\e)_{ij}$, are related by
mirror duality, which is a gauge transformation if $m\in G_L$.

\section{The $SU(3)$ Example}

Let us choose an orthogonal basis of the Cartan subalgebra
\be
H=\{H_1,H_2\},
\ee
and let $E_{\alpha}$ be the set of
generators corresponding to the six $SU(3)$ roots
\be
\a=\{\a_+,\a_-\}, \qquad \a_+=\{(\sqrt{3}/2,1/2),(0,1),
(-\sqrt{3}/2,1/2)\}, \qquad \a_-=-\a_+.
\ee
Here $\a_+$ ($\a_-$) are the positive (negative) roots, and $H_i, E_{\a}$
obey
\be
[H_i,H_j]=0,\qquad [H_i,E_{\a}]=\a_i E_{\a}, \qquad i,j=1,2 .
\ee

We now decompose the set of generators $T=\{H_i,E_{\a}\}$ into the direct
sum $T=T_+\oplus T_-$, where
\be
T_+ = \{H_+=H_1-iH_2,\, E_{\a_+}\}, \qquad T_- = \{H_-=H_1+iH_2,\, E_{\a_-}\};
\ee
these indeed obey the conditions of the theorem in section 2,
with a complex structure $h$ given in eq. (\ref{hTT}).
{}From eq. (\ref{mTT}) it follows that mirror transformation interchanges
$\a_+\leftrightarrow \a_-$, $H_-\leftrightarrow H_+$. Is it a gauge
transformation? The answer is yes, because the Weyl reflection that takes
$H_2\to -H_2$ (a reflection of the root (0,1)) does the job.

Therefore, mirror symmetry is a gauge symmetry in the $N=2$ moduli space
of the $SU(3)$ model (generated by adding deformations in the
Cartan); its action on the moduli space is induced by the transformation
$H_2\to -H_2$, as described in the previous section.

This example can be generalized to the $N=2$ moduli space of $A_{2n}$
models for all $n$;
In these cases mirror symmetry is a Weyl rotation, and therefore, it is a
gauge symmetry.

An example where mirror transformation is {\em not} a gauge symmetry is the
$N=2$, $SU(2)\times SU(2)$ model. In this case, in order to interchange the
positive roots with the negative roots by a Weyl transformation,
we need to reflect the Cartans of both $SU(2)$'s: $H_i\to -H_i, \; i=1,2$.
Such a transformation fails to interchange $H_+\leftrightarrow H_-$, and
therefore, it is not a mirror transformation.

\section{Conjectures: Mirror Duality as a Gauge Symmetry for Calabi-Yau
Compactifications}

We conclude with some speculations (which are the main motivation for this
work).
Although we have discussed mirror symmetry as a gauge symmetry in the
moduli space of $N=2$ backgrounds on a group $G$, we speculate that this
can be generalized to other examples (such as $N=2$ cosets
\cite{KS}~\footnote{
For Kazama-Suzuki models, $G/(H\times U(1))$, we can deform $(G/H)\times
U(1)$ to $G/(H\times U(1))\times U(1)^2$ at the boundaries, and duality
along the deformation line is a mirror transformation.}).
Mirror symmetry is particularly rich in the space of Calabi-Yau (CY)
compactifications \cite{mir}.
In what sense could mirror symmetry be a gauge symmetry for CY sigma-models?

At this stage, it is not known how to connect mirror pairs of CY
backgrounds by  marginal deformations of their corresponding
$c=9$ CFTs. This situation is similar to the
mirror pair of 2-$d$ `black hole' backgrounds
(the `cigar', $SL(2)/U(1)_a$, and the `trumpet', $SL(2)/U(1)_v$);
they cannot be connected by a marginal deformation of the $SL(2)/U(1)$ CFT.
But in the latter case we understand how to relate them by a mirror duality
which is a gauge symmetry: we look at the moduli space of 4-$d$, $N=2$
backgrounds connected to $SL(2)\times U(1)$. We then identify mirror
symmetry as a gauge symmetry in that moduli space and, in particular, at
the {\em boundary} it relates the cigar to the trumpet (times free
scalars).

The discussion above suggests that a mirror pair of CY backgrounds (times a
non-compact space) could
appear at the boundary of the moduli space of $d>6$, $N=2$ backgrounds.
Moreover, it might be possible that there is a {\em self-mirror} point
(with enhanced symmetry $G$) in the moduli space, and that mirror symmetry is a
gauge transformation in $G$.

Let us give some hints that this indeed could be true, at least for
particular CY backgrounds. Suppose we start with the $N=2$
$SU(2)_{k_1}/U(1)\times \prod_{i=2}^5 SU(2)_{k_i}$ model, such that the
total central charge is critical, $c=15$, and make a ``GSO projection'' by
twisting with exp$(2\pi iJ_0)$, where $J$ is the $N=2$ $U(1)$ current.
Now, we deform this model with the four current-current operators at the
Cartan sub-algebra of the four $SU(2)$'s, simultaneously. At the boundaries of
the deformation line, one gets the product of $\prod_{i=1}^5
SU(2)_{k_i}/U(1)$
(with central charge $c=9$) with non-compact $U(1)^4$ (with central charge
$c=6$). At one boundary it is twisted only by the GSO projection, and
the coset CFT $\prod_{i=1}^5 SU(2)_{k_i}/U(1)$ is related to the
CFT sigma-model on a CY manifold in ${\bf CP}^4$~\cite{Gepner}.
At the other boundary it can be viewed as being twisted by the product of
$Z_{k_i}$'s (in the same way it works for a single minimal model); combined
with the GSO projection it gives rise to the mirror manifold (when acting
on the CY sigma-model corresponding to the product of minimal models).

It should be mentioned that although the duality is not a mirror
transformation along the deformation line, it is a mirror transformation
acting on the $c=9$ CY background at the boundary (without acting on the
decoupled 4-$D$ non-compact flat space). In the sigma-model
description in terms of manifolds admitting $N=2$, it is therefore suggested
that mirror symmetry for CY backgrounds of that type is indeed a gauge
symmetry.

\vskip .3in \noindent
{\bf Acknowledgements} \vskip .2in \noindent
Research of E.W. is supported in part by NSF grant PHY92-45317.
Research of A.G. is supported in part by BSF - American-Israeli Bi-National
Science Foundation and by an Alon fellowship.

\end{document}